\documentclass[preprint,showpacs,preprintnumbers,amsmath,amssymb]{revtex4-1}
\usepackage{graphicx}
\usepackage{float}
\usepackage{color}
\usepackage[pdftex]{hyperref} 
\hypersetup{
colorlinks,%
citecolor=blue,%
filecolor=black,%
linkcolor=black,%
}
\usepackage{dcolumn}
\usepackage{bm,verbatim}
\usepackage{amsmath} 
\usepackage{bm,lineno} 
\usepackage{natbib} 
\graphicspath{E:\Work\Article 4 (Nov 2009) - Mode-selective vibrational excitation\Article\LaTex}
\begin{document}
\title{Mode-selective vibrational excitation induced by nonequilibrium
 transport processes in single-molecule junctions}
\author{Rainer H\"artle$^{(a)}$}
\author{Roie Volkovich$^{(b)}$}
\author{Michael Thoss$^{(a)}$}
\author{Uri Peskin$^{(b)}$}
\affiliation{$^{(a)}$Institut f\"ur Theoretische Physik and Interdisziplin\"ares Zentrum f\"ur Molekulare Materialien, Friedrich-Alexander-Universit\"at Erlangen-N\"urnberg,\\ 
Staudtstr.\,7/B2, D-91058 Erlangen, Germany\\
$^{(b)}$Schulich Faculty of Chemistry,\\
and the Lise Meitner Center for Computational Quantum Chemistry\\ 
Technion-Israel Institute of Technology, Haifa 32000, Israel}
\begin{abstract}
In a nanoscale molecular junction at finite bias voltage, 
the intra-molecular distribution of vibrational energy can strongly deviate
from the thermal equilibrium  distribution and specific vibrational modes can be
selectively
excited in a controllable way, 
regardless of the corresponding mode frequency. This is demonstrated for generic models of 
asymmetric molecular junctions with localized electronic states, employing a master equation as well as a nonequilibrium Green's function approach. 
It is shown that the applied bias voltage controls the  excitation
of specific vibrational modes
coupled to these states,
by tuning their electronic population, 
which influences the efficiency of vibrational cooling processes due to energy exchange with the leads.

\vskip 6 cm
uri@tx.technion.ac.il
\end{abstract}
\maketitle
Nonequilibrium processes in nanoscale quantum systems
and particularly quantum transport processes in molecular junctions \cite{9,10}
have received great attention recently. 
Coupling a single molecule to
two metal or semi-conductor electrodes at different electrochemical potentials
drives it into a nonequilibrium steady-state which can deviate significantly 
from thermal equilibrium. The mechanical stability of such a
junction and its transport properties strongly depend 
on the distribution of vibrational energy between the
different molecular modes. Experiments employing scanning tunneling microscopy
(STM)  \cite{18,21,22}
or molecular junctions \cite{23,ad1,ad4}
as well as theoretical studies \cite{ad2,ad3,25,42,29,30,31,33,50} 
indeed demonstrated nonequilibrium vibrational
excitation, as well as conformational 
changes and chemical reactions induced by charge transport under finite bias.

In this communication we demonstrate that the molecular junction scenario 
offers a new route to mode-selective
excitations of molecules. Mode-selective excitations were discussed 
extensively in the past in the context of photo-induced chemical reactions
\cite{1,3,4}. 
The original idea was that energy can be directed into specific
mode-excitation or bond-cleavage, prior to its statistical randomization due
to internal 
vibrational energy redistribution (IVR) processes. Since statistical distribution of
energy usually prevails in polyatomic molecules and 
in molecules in condensed phase environments, typically the lower frequency modes are
 excited and the weaker 
bonds are broken. The challenge of overcoming this statistical behavior led to
control schemes based on 
laser excitations \cite{6,7,8}. While the excitation of a molecule with a
laser pulse results in transient
nonequilibrium dynamics followed by relaxation to equilibrium, in a biased
molecular junction a 
stationary nonequilibrium state is maintained. 

The external bias voltage, as we will show, can be used to drive the molecule into
different nonequilibrium internal energy 
distributions, thus enabling voltage-controlled selective excitations of
specific vibrational motions, conformational 
changes or, in principle, bond dissociation. While differences in symmetry between different
vibrational modes lead to well known
selection and/or propensity rules \cite{34,35} for excitations, in this work
we focus on bias-controlled selective 
excitations induced by nonequilibrium effects and use a simple model to demonstrate how non-thermal
excitations of specific molecular modes can be 
independently controlled by the bias voltage.

\begin{figure}
\includegraphics[scale=0.38]{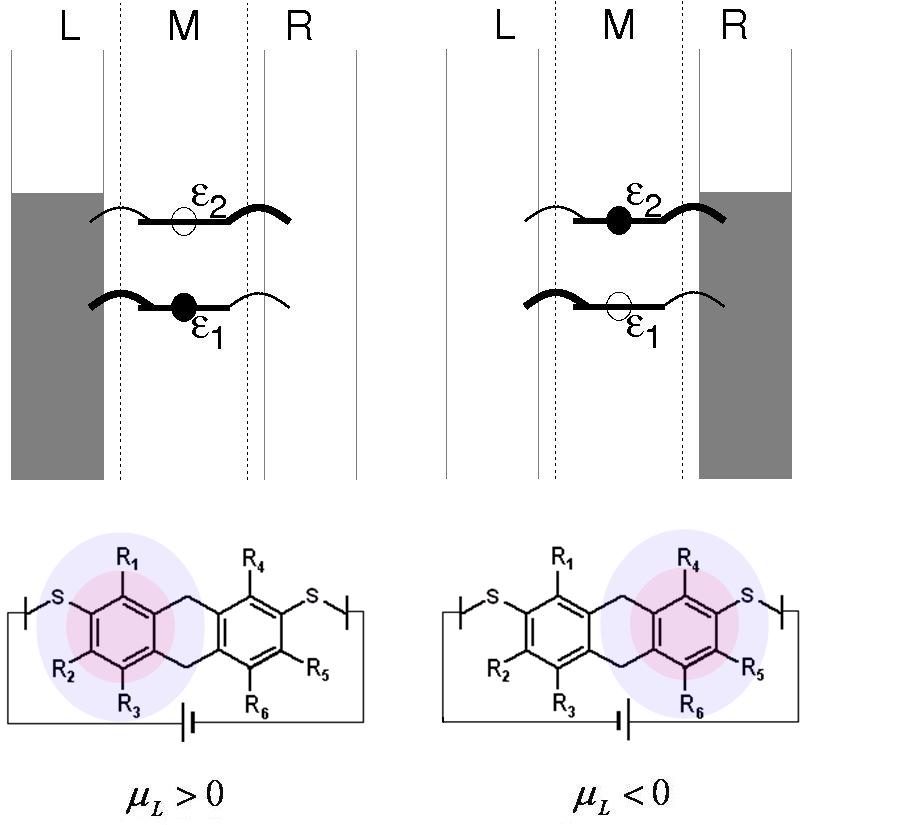}
\caption{\label{fig1}}
\end{figure}

To demonstrate and analyze the possibility of mode-selective vibrational excitation, we consider 
here a representative model for vibronic coupling in an asymmetric molecular junction. 
The model, schematically depicted in Fig.\ \ref{fig1},
comprises two electronic states on the molecular bridge, which are assumed to be localized at different parts of the molecule.
This localization can be achieved, \textit{e.g.}, 
employing electron donating or withdrawing functional side groups as illustrated in Fig.\ \ref{fig1}, 
and results in an asymmetry in the coupling of the electronic states to the leads such that the population of the electronic states can
be controlled by the bias direction (see below). Furthermore, due to the asymmetric charge distribution, 
the coupling of the electronic to the vibrational
degrees of freedom will be asymmetric as well, thus allowing for selective 
current-induced vibrational excitation.

The Hamiltonian describing vibrationally coupled electron transport in this model is given by
\begin{eqnarray}
H=H_\text{M}+H_{\text{Leads}}+H_{\text{M},\text{Leads}},
\end{eqnarray}
where the different parts correspond to the molecular bridge, the leads, and the molecule-lead coupling, respectively.
The molecular part of the Hamiltonian, $H_\text{M}=H_\text{0}+H_{\text{B}}$
with
\begin{eqnarray}\label{H0}
H_{0}&=&\sum^{N_{\text{el}}}_{m=1}\epsilon_ma_m^{\dagger}a_m+\sum^{N_{\text{vib}}}_{\nu=1}\Omega_{\nu}c_\nu^{\dagger}c_\nu 
+ \sum^{N_{\text{vib}}}_{\nu=1}\frac{c_\nu^{\dagger}+c_\nu}{\sqrt{2}}\sum^{N_{\text{el}}}_{m,m'=1}\lambda_{\nu,m,m'} a_m^{\dagger}a_{m'} ,
\end{eqnarray}
involves $N_{\text{el}}$  electronic states on the molecular bridge with energies $\epsilon_m$ and the corresponding creation and annihilation operators $a_m^{\dagger}$ and $a_m$.
The vibrational degrees of freedom of the molecular bridge ('internal modes') are described by harmonic modes with frequencies $\Omega_{\nu}$, corresponding creation and annihilation operators
$c_\nu^{\dagger}$ and $c_\nu$,
and are assumed to be linearly coupled to the 
electronic degrees of freedom. 
To incorporate vibrational relaxation effects, each internal vibrational mode is coupled to a thermal bath described by the Hamiltonian
\begin{eqnarray}
H_{\text{B}} &=&\sum^{N_{\text{vib}}}_{\nu=1}\sum^{N_{\text{bath}}}_{\beta_{\nu}=1}[{\omega_{\beta_{\nu}} d^\dagger_{\beta_{\nu}}}{d_{\beta_{\nu}}}
+(c_v^{\dagger}+c_v){\eta_{\nu,\beta_{\nu}}(d^\dagger_{\beta_{\nu}}}+d_{\beta_{\nu}})].
\end{eqnarray}
Here, $d^\dagger_{\beta_{\nu}}, d_{\beta_{\nu}}$ denote the creation and annihilation operator for a bath mode with frequency
$\omega_{\beta_{\nu}}$, and $\eta_{\nu,\beta_{\nu}}$ is the system-bath coupling strength.
All properties of the bath that influence the dynamics of the system are described by the spectral 
densities $J_\nu(\omega)= \sum^{N_{\text{bath}}}_{\beta_{\nu}=1}\eta_{\nu,\beta_{\nu}}^{2}\delta(\omega-\omega_{\beta_{\nu}})$. In the calculations 
presented below, we have assumed an Ohmic bath model,
\textit{i.e.} $J_\nu(\omega)= \frac{\zeta_\nu^2}
{\omega_{c,\nu}^2}\omega e^{-\omega/\omega_{c,\nu}}$.
 
The electrodes are modeled as non-interacting electron reservoirs with electrochemical potential $\mu_{\text{L/R}}$. 
The electronic states in the right and left electrodes are characterized by their
energies {$\epsilon_{k}$}, and the corresponding electron creation and annihilation 
operators $b^\dagger _{k}$ and $b_{k}$, resulting in the Hamiltonian
$H_{\text{Leads}}=\sum_{k\in \text{L,R}}\epsilon_k b_k^\dagger b_k $. 
Finally, the molecule-lead coupling is given by 
$H_{\text{M},\text{Leads}}=\sum_{k\in \text{L};m} \upsilon_{\text{L},m}\xi_{\text{L},k}a_m^{\dagger}b_k
+ \sum_{k\in \text{R},m} \upsilon_{\text{R},m}\xi_{\text{R},k}a_m^{\dagger} b_k+\textit{h.c.}$\,.
Here, $\upsilon_{\text{L/R},m}$ are dimensionless factors that determine the asymmetry in the 
coupling of the $m$th molecular state to the right or the left electrode, and the molecule-lead coupling strengths are defined as $\xi_{\text{L/R},k}$ and correspond to 
width functions $\Gamma_{\text{L/R}}(\epsilon)= 2\pi \sum_{k \in \text{L/R}}\vert\xi_{\text{L/R},k}\vert^{2}\delta(\epsilon-\epsilon_k)$.
In the calculation presented below the leads are modeled as semi-infinite tight-binding chains, where the width-function for the semielliptic band is $\Gamma_{\text{L/R}}(\epsilon)\equiv\frac{|\xi|^2}{|\gamma|^2}\sqrt{(2\vert\gamma\vert)^{2}-(\epsilon-\mu_{\text{L/R}})^2}$
with bandwidth  $4\gamma$ and overall molecule-lead coupling strength $\xi$.

Nonequilibrium transport in molecular junctions is studied employing two complementary approaches, 
where the technical details are summarized in the Supporting Information \cite{ad5} (see also Refs.\ \onlinecite{29,30}).
Briefly, the first approach is based on propagating in time
a reduced density matrix $\rho_\text{S}(t)$, which incorporates the electronic states of the molecular bridge,
the internal vibrational modes and the vibronic coupling in a non-perturbative manner \cite{rev}.
A master equation for $\rho_\text{S}(t)$ is obtained employing perturbation theory 
in the molecule-lead and system-bath coupling as well as the Markov approximation \cite{36,37,41}. 
Steady-state observables are determined at the infinite-time limit, \textit{e.g.}, the
internal mode excitation is given by
$ \langle c^\dagger_\nu c_\nu\rangle=\lim_{t \rightarrow \infty}\text{tr}_\text{S}[\rho_\text{S}(t)c^\dagger_\nu c_\nu]$.
The second approach is based on nonequilibrium Green's functions (NEGF). 
Specifically, we have applied a method originally proposed by Galperin \emph{et al.} \cite{42}, which was recently extended to 
account for multiple vibrational modes and multiple electronic states \cite{29,30}. 
This approach is complementary to the master equation approach, because it is non-Markovian
and non perturbative in the molecule lead coupling, while vibronic coupling is treated in an approximate non-perturbative way.
Thus, the NEGF method provides a more comprehensive description of the interactions between the molecule and the leads, while the master equation approach invokes no explicit approximations with respect to the electron-vibrational coupling. The good agreement of the results presented below demonstrates the validity of both approaches for the selected model parameters. The results also show that coherences between the molecular levels, which are not included in the master equation approach but in the NEGF scheme, are negligible. 

Various possible mechanisms can result in mode-selective vibrational excitation. In the following, we discuss mechanisms which are expected
to be particularly relevant in nanoscale molecular junctions. In all cases we consider the model described above 
with two electronic states, which are coupled asymmetrically to the two electrodes and selectively to two molecular modes, 
corresponding to
$\lambda_{\nu,m,m'}=\delta_{m,m'}\lambda_{\nu,m}$, with $\lambda_{1,2}=\lambda_{2,1}=0$ in
Eq.\ (\ref{H0}). 
The specific model parameters are detailed in Table \ref{table}. 
We note that the mode-selective vibrational excitation reported below is 
also obtained for models with small to moderate
off-diagonal vibrational coupling terms,
$\lambda_{1,2}=\lambda_{2,1}\lessapprox\lambda_{1,1}/10$, but becomes less
efficient for larger off-diagonal coupling strengths.

\begin{figure}
\begin{center}
\includegraphics[scale=0.4]{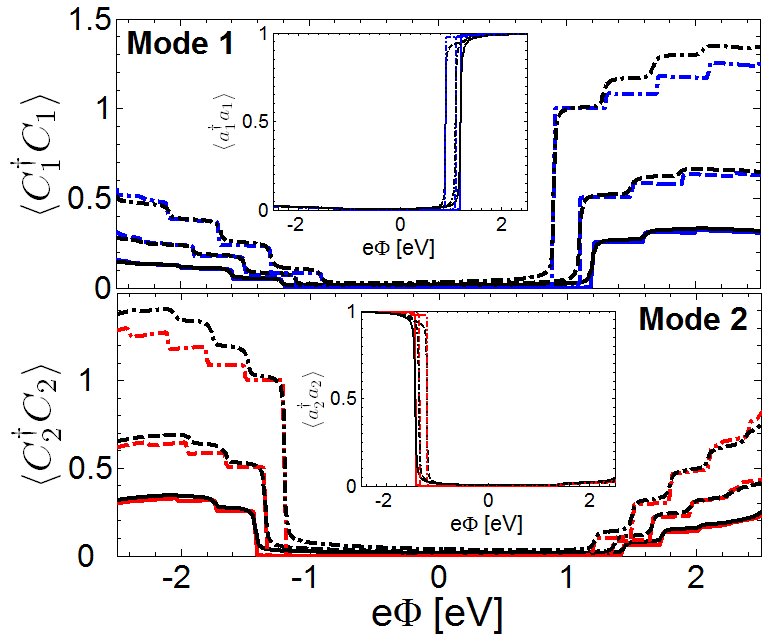}
\end{center}
\caption{\label{fig2}}
\end{figure}

Fig.\ \ref{fig2} shows results for the vibrational excitation of the two
internal modes as a function of the applied
bias voltage for a model with a moderate coupling of the
internal modes to a thermal bath. 
The results demonstrate that the vibrational excitation of each specific mode depends 
significantly on the bias voltage and can be controlled by its direction. 
This finding can be rationalized considering that 
the effective electronic-vibrational coupling (and thus the vibrational excitation)  
of a specific mode depends on the residence time
of the electrons on the molecule (relative to the vibrational period), and thus
on the occupation of the electronic state to which the mode is coupled.
The inset of Fig.\ \ref{fig2} shows that
the occupation of the two electronic states of the asymmetric junction (see Fig. 1) is sensitive 
to the bias direction, so that in each direction the unoccupied state is associated with 
lower vibrational excitation. 

\begin{figure}
\begin{center}
\includegraphics[scale=0.55]{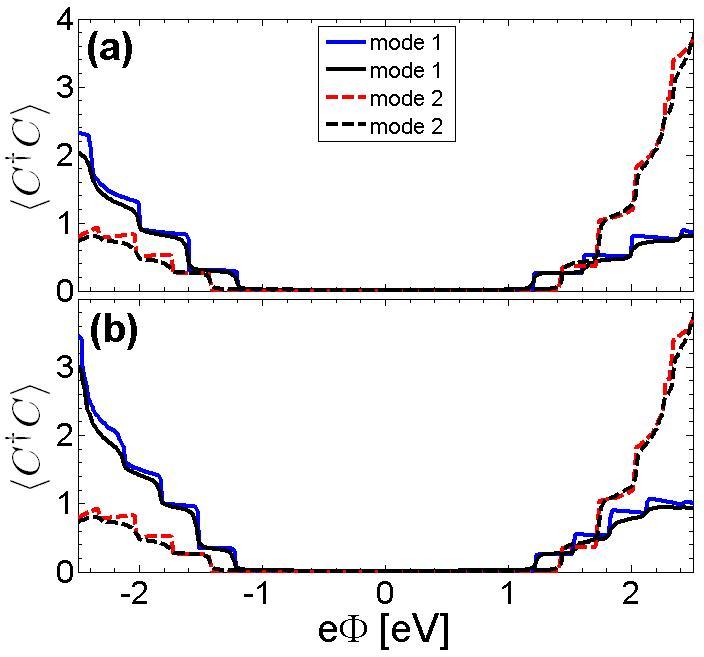}
\end{center}
\caption{\label{fig4}}
\end{figure}

\begin{figure}
\begin{center}
\includegraphics[scale=0.38]{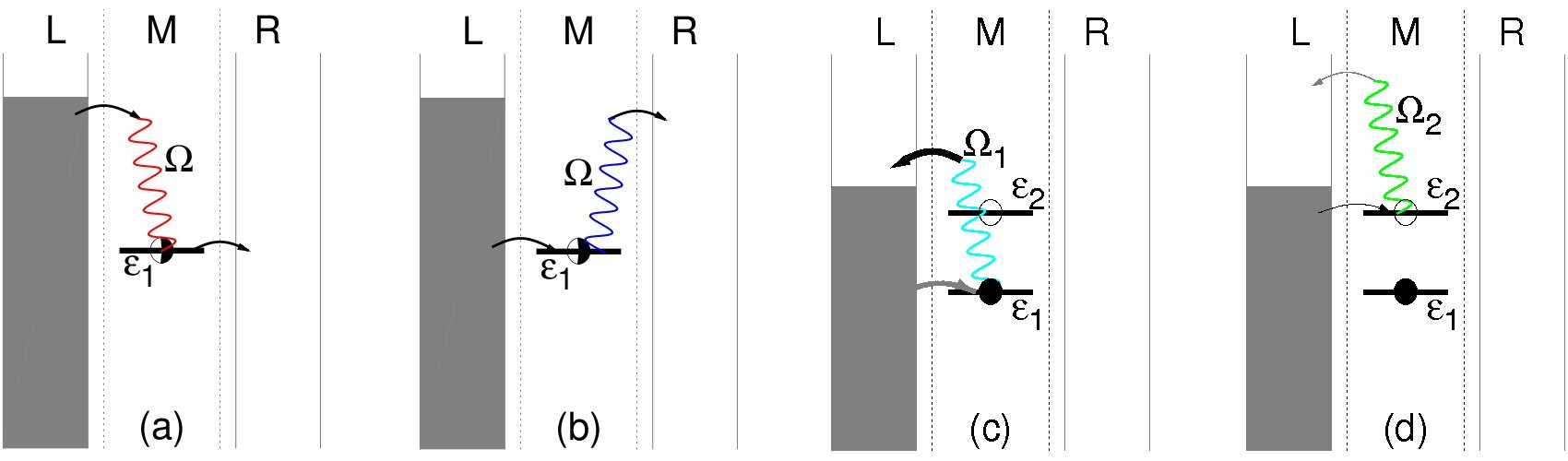}\\
\end{center}
\caption{\label{fig5} }
\end{figure}

The mechanism of current-induced vibrational excitation and the dependence on the bias
(direction) can be significantly more complex and intriguing if 
the cooling by the external thermal bath is less effective and other 
nonequilibrium effects prevail. As an example, Fig.\ \ref{fig4} depicts results for a model with a weak coupling to a thermal bath.
The results show that the overall vibrational excitation is
significantly larger, revealing
efficient current-induced heating processes schematically depicted in 
Fig.\ \ref{fig5}a.
The level of vibrational excitation for the two modes depends again significantly and
in a nontrivial way on the bias voltage. The results presented in Fig.\ 3b for a model where the 
two vibrational modes have identical frequencies, $\Omega_1=\Omega_2$, and the same vibronic
coupling strengths, $\lambda_{1,1}=\lambda_{2,2}$, reveal furthermore that this behavior is not solely caused 
by different vibrational excitation energies or vibronic coupling strengths but rather results 
from the asymmetry of the molecular junction. It is important to note that 
the relation between the degree of vibrational excitation of a given mode 
and the occupation of the electronic state, to which the mode is coupled, is reversed with respect 
to the results obtained above with a significant 
coupling to a thermal bath \cite{43}
(cf.\ Fig.\ \ref{fig2}), 
\textit{i.e.}, vibrational excitation is less significant for the mode 
coupled to the occupied electronic state. This intriguing finding suggests that other 
cooling mechanisms are active that are selectively associated with the 
occupied molecular level. 

Indeed, in the absence of efficient cooling by a thermal bath the vibrational excitation of 
the internal modes due to "heating" processes can be limited only by energy 
loss to the electrons in the leads. One of these cooling mechanisms is provided 
by the transport process itself, where upon electron transmission from one lead to another a 
vibrational quantum is absorbed (Fig.\ \ref{fig5}b), instead of being emitted (Fig.\ \ref{fig5}a).
Notice that when the non-equilibrium vibrational distribution involves significant population 
of excited states, cooling by phonon absorption becomes as efficient as heating by phonon emission. However, 
since transport processes depend on the molecular coupling to both of the leads, they are not 
expected to distinguish between the two 
electronic states or to selectively influence the 
respective mode excitation. In contrast, cooling processes that involve only one lead, 
directly transfer the junction asymmetry to the vibrational system. 
Two such processes, which involve the creation of an electron-hole pair in the left lead \cite{250} by absorbing a single vibrational 
quantum \cite{22}, are shown in Fig.\ \ref{fig5}c and Fig.\ \ref{fig5}d. 
When the applied bias is sufficiently large, $\Phi\gtrapprox2\epsilon_{1/2}$, the 
asymmetry in the coupling of state 1 and state 2 to the left lead (sketched by the 
thickness of the transmission arrows in Fig.\ \ref{fig5}c and Fig.\ \ref{fig5}d) 
implies that most of the electron-hole pair creation processes in the left lead 
involve state 1. As a result of these absorption processes, the mode coupled
to state 1 looses significantly more energy to the left lead than the mode coupled to state 2.
Similar electron-hole pair creation processes can occur in the right lead as well. 
However, provided that the two molecular states are located sufficiently above the 
Fermi energy of the electrodes, electron-hole pair creation in the right lead requires
absorption of multiple vibrational quanta and is thus less efficient. The net result 
is that for positive bias voltages electron-hole pair creation processes in the left 
electrode provide an exclusive source of cooling for the mode that is coupled to state 1, 
which is the occupied electronic state. Inverting the bias direction, cooling at the 
right lead becomes dominant, and again, the mode coupled to the occupied
electronic state (mode 2 in this case) is more effectively cooled. The bias direction therefore
controls the efficiency of the prominent vibrational cooling mechanism in the junction, 
and selectively controls the degree of mode excitation.
It is noted that this selective vibrational excitation depends on the two electronic states being located 
sufficiently above the Fermi energy of the electrodes, and thus can be switched off by shifting them 
towards the Fermi energy. One therefore has two possibilities to control 
this mechanism for mode-selective excitation: the direction of the bias voltage, which selects a specific mode, and the gate voltage,
which controls the degree of excitation of the selected mode. 

In summary, we have demonstrated and analyzed mode-selective vibrational excitation in
single-molecule junctions induced by nonequilibrium electron transport. 
The results obtained here for a generic model
show that necessary prerequisites for mode-selectivity in molecular junctions 
involve asymmetry of the molecule-lead coupling and/or the charge distribution
as well as state-dependent vibronic coupling. This can be controlled in a 
molecular junction by choosing
appropriate functional and/or anchor groups. The ability to use the external bias 
in order to direct energy selectively into a given molecular vibrational mode 
suggests the possibility of mode-selective chemistry, \textit{e.g.} the cleavage of
a specific, not necessarily the weakest, chemical bond induced by nonequilibrium transport processes.

{\bf Acknowledgments}

This research was supported by the German-Israeli Foundation for scientific development (GIF). 


\pagebreak
{\bf Figure Captions}

Fig.1: Sketch of the model molecular junction for positive ($\mu_{\text{L}}>0$) and negative  ($\mu_{\text{L}}<0$) bias voltages. The top figures depict the population of electronic levels. Occupied levels in the left (L) and the right (R) lead are represented by the gray-shaded areas. The population of the two electronic states of the molecular bridge, which are represented by two black bars, is indicated by filled and empty circles attached to the respective bars, respectively. Strong (weak) coupling of an electronic state to a lead is indicated by a thick (thin) arched line. 
The bottom figures give a schematic proposition of a scenario for such a molecular junction. 
Electron withdrawing ($R_1,R_2,R_3$) and electron donating ($R_4,R_5,R_6$)
groups attached to two phenyl rings should induce localization and asymmetric
charge distribution of the electronic
states, leading
to both bias controlled population of the molecular states and bias controlled excitation of the vibrational modes.

Fig.2: Voltage-dependent steady state vibrational excitation, 
demonstrating mode-selective vibrational excitation in the presence of a moderate coupling to a "cold"
bath ($k_{\text{B}}T=1m$eV). The solid, dashed and dot-dashed lines correspond to increasing vibronic coupling. The insets show the respective population of the electronic states. 
The model parameters are summarized in Table I. The black and colored curves are results of
NEGF and density matrix calculations, respectively.

Fig.3 Voltage-dependent mode-selective vibrational excitation
in a model with weak coupling to 
a bath, where nonequilibrium effects prevail. The black and colored curves are results of NEGF and density matrix calculations, 
respectively. The results have been obtained for a model with two vibrational
modes with different (a) and identical (b) frequencies and vibronic coupling
strengths, respectively (see Table I for the parameters values).

Fig. 4: Mechanisms of electronic-vibrational energy exchange in
molecular junctions. Mechanisms (a) and (b) 
represent heating and cooling induced by electronic transport between the two 
leads, respectively. Mechanisms (c) and (d), on the other hand, correspond to
cooling by electron-hole pair creation at the left electrode. 
The thin arrows in (d) reflect the inefficiency of the process due to weak coupling of the 
unoccupied state to the left lead at positive bias voltages.


\begin{table}[e] 
\caption{\label{Table I}Model Parameters (Energy values are in eV)}
\begin{ruledtabular}
\begin{tabular}{*{9}{c}}
$\epsilon_1$ & $\epsilon_2$ & $\upsilon_{L,1},\upsilon_{R,2}$ & $\upsilon_{R,1},\upsilon_{L,2}$ & $\omega_{c,1/2}$ & $\xi$ &  $\gamma$ & $\Gamma_{\text{L/R}}(\mu_{\text{L/R}})$ & $k_\text{B}T$ \\
\hline
0.65 & 0.75 & 1 & 0.1 & 1 & -0.1 & -1 & 0.02 & 0.001 \\
\end{tabular}
\end{ruledtabular}
\vskip 0.1 cm 
\begin{ruledtabular}
\begin{tabular}{*{6}{c}}
& FIG. 2 & FIG. 3a & FIG. 3b\\
\hline
$\Omega_1$ & 0.2 & 0.2 & 0.15 \\
$\Omega_2$ & 0.15 & 0.15 & 0.15 \\
$\frac{\lambda_{1,1}}{\Omega_1},\frac{\lambda_{2,2}}{\Omega_2}$ & 1/$\sqrt{2}$,1,$\sqrt{2}$ &1/$\sqrt{2}$ & 1/$\sqrt{2}$ \\
$\zeta_1,\zeta_2$ &-0.03 & -0.01 & -0.01\\
\end{tabular}
\end{ruledtabular}
\label{table}
\end{table}


{\bf References}

\begin{thebibliography}{99}
\bibitem{9} C. Joachim and S. Roth, \textit{Atomic and molecular wires} (Kluwer, Dordrecht, 1997). 
\bibitem{10} A. Nitzan and M. A. Ratner, Science 300, 1384 (2003). 
\bibitem{18} J. I. Pascual \emph{et al.}, Nature 423, 525 (2003). 
\bibitem{21} B.-Y. Choi \emph{et al.}, Phys. Rev. Lett. 96, 156106 (2006). 
\bibitem{22} G. Schulze \emph{et al.}, Phys. Rev. Lett. 100, 136801 (2008). 
\bibitem{23} Z. Ioffe \emph{et al.}, Nature Nanotech. 3, 727 (2008).
\bibitem{ad1} H. Park \emph{et al.}, Nature 407, 57-60 (2000).
\bibitem{ad4} H. Song \emph{et al.}, Appl. Phys. Lett. 94, 103110 (2009)) 
\bibitem{ad2} A. Mitra \emph{et al.}, Phys. Rev. B 69, 245302 (2004).
\bibitem{ad3} J. Koch \emph{et al.}, Phys. Rev. B 73, 155306 (2006).
\bibitem{25} J. Koch and F. von Oppen, Phys. Rev. Lett. 94, 206804 (2005).
\bibitem{42} M. Galperin \emph{et al.}, Phys. Rev. B 73, 045314 (2006). 
\bibitem{29} R. H\"artle \emph{et al.}, Phys. Rev. B 77, 205314 (2008). 
\bibitem{30} R. H\"artle \emph{et al.}, Phys. Rev. Lett. 102, 146801 (2009). 
\bibitem{31} V. May and O. K\"uhn, Chem. Phys. Lett. 420, 192 (2006). 
\bibitem{33} C-C. Kaun and T. Seideman, Phys. Rev. Lett. 94, 226801 (2005). 
\bibitem{50} A. Hackl \emph{et al.}, Phys. Rev. Lett. 102, 196601 (2009)
\bibitem{1} J. Jortner \emph{et al.}, \textit{Mode selective chemistry}, (Kluwer, Amsterdam, 1991).
\bibitem{3} Z. Liu \emph{et al.}, Science 312, 1024 (2006).
\bibitem{4} F. F. Crim, P. N. A. S. USA 105, 12654 (2008). 
\bibitem{6} E. D. Potter \emph{et al.}, Nature 355, 66 (1992). 
\bibitem{7} R. N. Zare, Science 279, 1875 (1998). 
\bibitem{8} P. W. Brumer and M. Shapiro, \textit{Principles of the Quantum Control of Molecular Processes}, (Wiley, 2003). 
\bibitem{34} A. Troisi and M. A. Ratner, Nano Lett. 6, 1784 (2006).
\bibitem{35} A. Gagliardi \emph{et al.}, Phys. Rev. B. 75, 174306 (2007). 
\bibitem{ad5} See Supplementary Material Document No for details.
\bibitem{rev} U. Peskin, J. Phys. B: At. Mol. Opt. Phys. 43 (2010).  
\bibitem{36} D. Egorova \emph{et al.}, J. Chem. Phys. 119, 2761 (2003). 
\bibitem{37} R. Volkovich \emph{et al.}, J. Chem. Phys. 129, 034501 (2008). 
\bibitem{41} E. G. Petrov and P. H\"anggi, Phys. Rev. Lett. 86, 2862 (2001).
\bibitem{43} Except for the small bias region, in which one of the molecular conductance channels is closed
\bibitem{250} A different mechanism for electron-hole pair creation, which involves an electronically excited state of the junction instead of a vibrationally excited state, is discussed \textit{e.g.} in Ref.\ \cite{251}.
\bibitem{251} B. D. Fainberg and M. Jouravlev and A. Nitzan, Phys. Rev. B 76, 245329 (2007)




\end{thebibliography}
\end{document}



\section{Supporting Information}

In this Supporting Information we briefly describe the two methods that we are using to compute transport properties of a molecular junction. 
We employ a time-dependent reduced density matrix approach (RDM) as well as a time-independent nonequilibrium Green's function approach (NEGF). In the following, we assume the vibrational couplings $\lambda_{\nu,m,m'}$ to be diagonal, \textit{i.e.} $\lambda_{\nu,m,m'}=\delta_{\nu,m}\delta_{m,m'}\lambda_{\nu,m}$.

\subsection{Reduced Density Matrix Approach}

This approach is based on the system's density matrix $\rho(t)\cong\rho_\text{S}(t)\otimes\rho_\text{B}^{\text{eq}}(t)$. Here, $\rho_\text{S}(t)$ denotes the reduced density matrix of the system (S) and $\rho_{\text{B}}^{\text{eq}}$ comprises the 'bath' degrees of freedom (B), namely the external nuclear modes and the electronic states in the left and the right lead. 
$\rho_\text{S}(0)$ can be chosen flexibly, and employing a second order expansion in the system-bath coupling strengths, its time-evolution follows a (Markovian) equation of motion  \cite{1,2,3,4}
\begin{eqnarray}
\frac{d\rho_\text{S}(t)}{dt}&=&-\frac{i}{\hbar}[H_\text{S},\rho_\text{S}(t)]-\frac{1}{\hbar^2}\int_0^t{\text{tr}_{\text{B}}{[H_{\text{SB}},[H_{\text{SB}}(\tau),\rho_\text{S}(t)\rho_\text{B}^{\text{eq}}
]]}d\tau}.
\end{eqnarray}
Here, the trace is taken over the degrees of freedom of the environment (external nuclear modes and electronic degrees of freedom of the leads) that are coupled to the system via the term $H_{\text{SB}}=H_{\text{M},\text{Leads}}+H_{\text{B}}-\sum^{N_{\text{vib}}}_{\nu=1}\sum^{N_{\text{bath}}}_{\beta_{\nu}=1}{\omega_{\beta_{\nu}} d^\dagger_{\beta_{\nu}}}{d_{\beta_{\nu}}}$ with $H_{\text{SB}}(\tau)=e^{-\frac{i}{\hbar}[H-H_{\text{SB}}]\tau}H_{\text{SB}}e^{\frac{i}{\hbar}[H-H_{\text{SB}}]\tau}$. Thereby, the bath density $\rho_{\text{B}}^{\text{eq}}(t)$ in the second order term 
is replaced by the equilibrium bath density $\rho_{\text{B}}^{\text{eq}}$.
Stationary nonequilibrium observables are defined by the infinite time limit of  $\rho_\text{S}(t)$,  where in particular, the average vibrational excitation number is given by
\begin{eqnarray}
\langle c^\dagger_\nu c_\nu\rangle=\lim_{t \rightarrow \infty}\text{tr}_\text{S}[\rho_\text{S}(t)c^\dagger_\nu c_\nu].
\end{eqnarray}
In steady-state, and for non-degenerate levels, it is sufficient to follow the populations of the vibronic eigenstates of the system (the diagonal elements of $\rho_\text{S}(t)$  \cite{5,6,7,8}), according to the matrix equation $\bm{\dot{\rho}}_\text{S}(t)=[\bm{\kappa}^{(\text{R})}+\bm{\kappa}^{(\text{L})}+\sum_\nu{\bm{\kappa}^{(\nu)}}]{\bm{\rho}}_\text{S}(t)$. State-to-state rates for electron exchange with the electrodes are thus determined by
\begin{eqnarray}
\label{megamma}
 [\kappa^{(\text{L/R})}]_{l,l'}&=&\frac{1-\delta_{l,l'}}{\hbar}(\Gamma^{\text{L/R};h}_{l,l'}+\Gamma^{\text{L/R};e}_{l,l'}) -\frac{\delta_{l,l'}}{\hbar}\sum_{l'\neq l}(\Gamma^{\text{L/R};h}_{l,l'}+\Gamma^{\text{L/R};e}_{l,l'})
\end{eqnarray}
with $\Gamma^{\text{L/R};\text{e/h}}_{l,l'}=\sum_m\vert\upsilon_{\text{L/R},m}\left[a^{\dagger}_{m}\right]_{l,l'}\vert^2 \Gamma_{\text{L/R}}(\epsilon_{l'}-\epsilon_{l})f_{\text{e/h}}^{\text{L/R}}(\epsilon_{l'}-\epsilon_{l})$,  and the respective electrons (e) and holes (h) Fermi distribution functions $f_{\text{e/h}}^{\text{L/R}}(\epsilon)$. Similarly, the state-to-state rates for phonon exchange with the nuclear bath
\begin{eqnarray}
\label{mespec}
 [\kappa^{(\nu)}]_{l,l'}&=&\frac{1-\delta_{l,l'}}{\hbar}(\Gamma^{\nu;d}_{l,l'}+\Gamma^{\nu;u}_{l,l'}) -\frac{\delta_{l,l'}}{\hbar}\sum_{l'\neq l}^N(\Gamma^{\nu;d}_{l,l'}+\Gamma^{\nu;u}_{l,l'})
\end{eqnarray}
are determined by the baths spectral densities and the phonon occupation numbers, where  $\Gamma_{l,l'}^{\nu;u/d}=2\pi\vert[c_\nu^\dagger+c_\nu]_{l',l}\vert^2 J_\nu(\epsilon_{l'}-\epsilon_{l})n_{u/d}(\epsilon_{l'}-\epsilon_{l})$,   $n_u(\epsilon)=\left(e^{\epsilon/k_\text{B}T}-1\right)^{-1}$, and $n_d(\epsilon)=n_u(\epsilon)+1$.

\subsection{Nonequilibrium Green's Function Approach}

An alternative approach is based on nonequilibrium Green's functions. Here, we apply the method originally proposed by Galperin \emph{et al.}\cite{9}, which was recently extended to account for multiple vibrational modes and multiple electronic states \cite{10,5}. The approach is based on the small polaron transformation of the Hamiltonian, $H\rightarrow\bar{H}$. The transformed Hamiltonian $\bar{H}$  comprises an exactly solvable part, where no electron-vibrational coupling term is present, and a renormalized molecule-lead coupling term, $\bar{H}_{\text{M},\text{Leads}}=\sum_m a^\dagger_m X_m^\dagger \left(\sum_{k\in \text{L}}{\upsilon_{\text{L},m}\xi_{\text{L},k}b_k}+\sum_{k\in \text{R}}{\upsilon_{\text{R},m}\xi_{\text{R},k}b_k}\right)+\textit{h.c.}$, where $X_m=\text{exp}\left({\sum_\nu \frac{\lambda_{\nu,m}(c_\nu-c_\nu^\dagger)}{\sqrt2\Omega_\nu}}\right)$.
We employ the following ansatz for the Green's function ${\ll
a_m(\tau)a_{m'}^\dagger(\tau') \gg_H}$:
\begin{eqnarray}
\label{decoupling}
 \ll a_m(\tau)a_{m'}^\dagger(\tau') \gg_H&=&\ll a_m(\tau)X_m(\tau)a_{m'}^\dagger(\tau') X^\dagger_{m'}(\tau')\gg_{\bar{H}}\nonumber\\
&\approx&\ll a_m(\tau)a_{m'}^\dagger(\tau') \gg_{\bar{H}}\ll X_m(\tau)X_{m'}^\dagger(\tau') \gg_{\bar{H}},
\end{eqnarray}
which is written as a product of an electronic Green's function $\bar{G}_{m,m'}(\tau,\tau')= {\ll a_m(\tau)a_{m'}^\dagger(\tau') \gg_{\bar{H}}}$ and a correlation function of the shift operators $X_{m}$ and $X^\dagger_{m'}$. Thereby, the indices $H/\bar{H}$  denote the Hamiltonian, by which the respective contour-ordered correlation functions are evaluated. Employing an equation of motion approach, the self-energy matrices for the electronic part can be determined to second order in the molecule-lead coupling
\begin{eqnarray}
\Sigma_{m,m',\text{L/R}}(\tau,\tau')=\sum_{k\in \text{L/R}}\upsilon_{\text{L/R},m}\upsilon_{\text{L/R},m'}^*|\xi_{\text{L/R},k}|^2 g_k(\tau,\tau')\ll X_{m'}(\tau')X_{m}^\dagger(\tau) \gg_{\bar{H}},
\end{eqnarray}
where {$g_k(\tau,\tau')$} denotes the free Green's function of lead state $k$. \\
Defining the momentum correlation functions ${ D_{\nu,\nu'}=(i/2)\ll (c_\nu(\tau)-c_\nu^\dagger(\tau))(c_{\nu'}(\tau')-c_{\nu'}^\dagger(\tau'))\gg_{\bar{H}} }$, the shift operators correlation functions can be obtained using a cumulant expansion in the dimensionless coupling parameters $\lambda_{\nu,m}/\sqrt2\Omega_\nu$:
\begin{eqnarray}
\ll X_m(\tau)X_{m'}^\dagger(\tau')
\gg_{\bar{H}}=\text{exp}\left(\sum_{\nu,\nu'}i\frac{\lambda_{\nu,m}\lambda_{\nu',m'}}{\Omega_\nu
\Omega_{\nu'}}D_{\nu,\nu'}(\tau,\tau')-i\frac{\lambda_{\nu,m}^2+\lambda_{\nu',m'}^2}{2\Omega_\nu
\Omega_{\nu'}}D_{\nu,\nu'}(\tau,\tau)\right).
\end{eqnarray}
The vibrational self-energy matrix $\Pi_{\nu,\nu'}$, which represents the self-energy matrix for $D_{\nu\nu'}$, comprises two parts: $\Pi_{\text{bath},\nu,\nu'}$, which describes the coupling of the internal to the external modes, and
\begin{eqnarray}
\label{Piel}
\Pi_{\text{el},\nu,\nu'}(\tau,\tau')=-i\sum_{m,m'}\frac{\lambda_{\nu,m}\lambda_{\nu',m'}}{\Omega_\nu \Omega_{\nu'}}(\Sigma_{m,m'}(\tau,\tau')\bar{G}_{m',m}(\tau',\tau)+\Sigma_{m',m}(\tau',\tau)\bar{G}_{m,m'}(\tau,\tau')), 
\end{eqnarray}
which accounts for the interactions between the internal modes and the electrons. Since $\Pi_{\nu,\nu'}$  depends on the electronic self-energies $\Sigma_{m,m'}=\Sigma_{m,m',\text{L}}+\Sigma_{m,m',\text{R}}$ and Green's functions $\bar{G}_{m,m'}$, the respective equations need to be solved iteratively in a self-consistent scheme.\\
Nonequilibrium transport properties are expressed in terms of the Green's functions and self-energies, where \textit{e.g.} the vibrational excitation to second order in the coupling between the internal and the bath modes reads  \cite{9}: 
 \begin{eqnarray}
<c^\dagger_\nu c_\nu>_H &=&-\left(A_{\nu}+\frac{1}{2}\right)\text{Im}\left[D^{<}_{\nu,\nu}(t=0)\right]-\left(B_{\nu}+\frac{1}{2}\right)+\sum_{m}\frac{\lambda_{\nu,m}^{2}}{\Omega_\nu^2}\text{Im}[\bar{G}^<_{m,m}(t=0)], \nonumber\\ 
&&\hspace{-1cm}A_{\nu}=\mathcal{P}\sum_{\beta_{\nu}} \frac{\vert \eta_{\nu,\beta_{\nu}}\vert^{2}\omega_{\beta_{\nu}}}{\Omega_{\nu}(\omega_{\beta_{\nu}}^{2}-\Omega_{\nu}^{2})},\quad B_{\nu}=\mathcal{P}\sum_{\beta_{\nu}} \frac{\vert \eta_{\nu,\beta_{\nu}}\vert^{2}}{\omega_{\beta_{\nu}}^{2}-\Omega_{\nu}^{2}}\left(1+2n_{u}(\omega_{\beta_{\nu}})\right).\nonumber
\end{eqnarray}

\subsection{Discussion}

Both approaches are designed for molecular junctions with a weak coupling between the molecular bridge and the leads. Different approximations and assumptions are inherent to both methods, though.
For example the existence of a steady state transport regime has to be assumed within the NEGF formalism, while it is an outcome of the time evolution within the RDM method, where it is explicitly determined from a flexibly chosen initial state $\rho_{\text{S}}(t=0)$. Furthermore, although both approaches 
describe all resonant transport processes, the RDM method misses high order processes like co-tunneling due to a strict second order expansion of the (Markovian) equation of motion for $\rho_{\text{S}}(t)$ in the molecule-lead coupling. These processes, which result in a broadening of the resonances in the transport characteristics, are accounted for by the NEGF formalism that includes higher order processes in terms of Dyson series. 
Moreover, the present formulation of the RDM approach involves just the level-width functions $\Gamma_{\text{L/R}}$ and the spectral densities $J_\nu$ (cf.\ Eqs.\ (\ref{megamma}) and (\ref{mespec})), and thus, 
does not account for
the renormalization of the molecular energy levels due to the system-bath coupling. 
The NEGF approach does account for both the renormalization of the energy levels and coherences, which can be crucial for the description of degenerate levels. In contrast to the NEGF approach that is based on the decoupling approximation Eq.\ (\ref{decoupling}) as well as the noncrossing approximation\cite{9,10}, which is employed for Eq.\ (\ref{Piel}), the RDM approach is capable of describing all interactions on the molecular bridge exactly.\\
Therefore, to some extent, both approaches complement each other, and overall, we find a reasonably good agreement of the results obtained with the two methods, which we interpret as a strong indication of their validity for the selected model parameters.